# Nanoscale spatially resolved mapping of uranium enrichment in actinide-bearing materials

Elizabeth Kautz[a], Douglas Burkes[a], Vineet Joshi[b], Curt Lavender[b], Arun Devaraj[c]*

[a]National Security Directorate, Pacific Northwest National Laboratory, 902 Battelle Boulevard, P.O. Box 999, Richland, WA 99352, United States

[b]Energy and Environment Directorate, Pacific Northwest National Laboratory, 902 Battelle Boulevard, P.O. Box 999, Richland, WA 99352, United States

[c]Physical and Computational Sciences Directorate, Pacific Northwest National Laboratory, 902 Battelle Boulevard, P.O. Box 999, Richland, WA 99352, United States

*corresponding author: arun.devaraj@pnnl.gov

**ABSTRACT**

Spatially resolved analysis of uranium isotopes in small volumes of actinide-bearing materials is critical for a variety of technical disciplines, including earth and planetary sciences, environmental monitoring, bioremediation, and the nuclear fuel cycle. However, achieving sub-nanometer scale spatial resolution for such isotopic analysis is currently a challenge. By using atom probe tomography, a three dimensional nanoscale characterization technique, we demonstrate unprecedented nanoscale mapping of uranium isotopic enrichment with high sensitivity across various microstructural interfaces within small volumes (100 nm$^3$) of depleted and low enriched uranium alloyed with 10 wt % molybdenum with different nominal enrichments of 0.20 and 19.75% $^{235}$U respectively. The approach presented here can be applied to study nanoscale variations of isotopic abundances in the broad class of actinide-bearing materials, providing unique insights into their origin and thermo-mechanical processing routes.

Key words: uranium enrichment, atom probe tomography, low-enriched uranium, nuclear fuel, U-Mo, isotopic abundance, actinide-bearing material

Uranium (U) is the heaviest element naturally occurring in the Earth's crust in significant amounts, and is used both in its natural, processed, and anthropogenic forms. U isotopes are central to a diverse set of scientific disciplines, most notably earth and planetary sciences [1-8], toxicology, environmental monitoring and bioremediation [9-11], and the nuclear fuel cycle, forensics, and safeguards [12-22]. Specifically in nuclear fuel cycle applications, the amount of the fissionable U isotope ($^{235}$U) relative to all U (defined as enrichment) is a critical parameter in fuel performance, and thus impacts economic viability of nuclear power. Increasing U enrichment beyond natural abundance of 0.7% is necessary in order to allow for a self-sustaining fission chain reaction to proceed in a nuclear fuel [23]. The distribution of $^{235}$U in a nuclear fuel before irradiation can directly influence the nucleation and distribution of fission products and fission gas bubbles leading to impact on fuel swelling kinetics [24]. Fuel swelling in turn impacts geometric stability (e.g. bowing, deflection), robustness of fuel plates under irradiation, impacting phenomenon such as coolant channel closure, or release or fission gas products to the coolant [25]. These phenomena make it critical to analyze distribution of $^{235}$U in a nuclear fuel at a high spatial resolution to account for $^{235}$U enrichment variation across all possible nanoscale heterogeneities in the microstructure. The small volume (nanometers to micrometers in diameter) of precipitates or interfacial regions that must be analyzed limits the ability of many bulk analysis techniques currently used for quantifying $^{235}$U enrichment, introducing a crucial technological and resultant knowledge gap. Additionally, the risk associated with handling radioactive materials is high, and analysis of small volumes is needed to minimize risk to researcher health and the spread of radioactive contamination. Our work aims to address this gap, while also demonstrating a methodology by which $^{235}$U isotopic abundances in actinide-bearing materials can be measured quantitatively with sub-nanometer scale spatial resolution to gain uniquely powerful insight into material radioactivity, origin, or processing history.

As a part of U.S. high performance research reactor conversion program, there is an effort to replace all the highly enriched uranium fuel currently used in research reactors to low enriched nuclear fuels. Development of such a LEU fuel has significant implications to international nuclear non-proliferation, safeguards, and health and environmental contamination risks associated with continued handling of highly enriched uranium (HEU) fuels [18, 25-30]. Metallic low enriched fuel made of Uranium-10wt% Molybdenum alloys (LEU-10Mo) with less than 20% enrichment is a leading candidate chosen for this effort. As a part of fuel qualification and testing efforts, multiple batches of various LEU fuel plates (e.g. U-Mo, U-Si, U-Al) have been fabricated and irradiated in research reactors and several studies examined the nuclear fuel microstructure after neutron irradiation to improve understanding of material behavior in a reactor environment [25, 31, 32]. Microstructural features observable in fuels after irradiation via optical and electron microscopy techniques range from large fission gas bubbles to inclusions and elemental segregation. To develop a comprehensive atomic-level understanding of fuel performance in reactor, it is critical to understand the distribution of the fissionable $^{235}$U isotope in the starting fuel microstructure, at a nanoscale spatial resolution to determine the isotopic distribution in different phases and across interfaces in the microstructure.

Measurement of U enrichment in small volumes of actinide-bearing materials at nanoscale spatial resolution is also crucial to a variety of other fields. For example, environmental monitoring and bioremediation efforts in the wake of nuclear accidents (such as Fukushima Daiichi, Three Mile Island, or Chernobyl) and detonation of nuclear weapons (for testing or in war time) crucially depends on detection and mapping of actinide speciea. As part of the environmental monitoring process, measurement of U concentration in geological materials (soil, sandstone ores, and other organics materials) or particulate matter (produced from a detonation event) is performed. The measurement of U isotopic abundances can allow for improved understanding of atomic-scale transport of radionuclides in natural materials, the origin of the particles (including age and processing history), how radioactive particles spread, and how harmful they are to the surrounding environment and communities [9, 10, 22]. Another scientific area where U isotopic enrichment measurement is critical is in regulating the international transportation of actinide-bearing materials. Isotopic measurements in nuclear materials and comparison to existing models is necessary for understanding sample history, and ensuring safe handling of nuclear materials [14, 21, 22]. High spatial resolution analysis of U enrichment of small volumes can critically impact these wide variety of scientific areas.

Current capabilities for measurement of U isotope abundances typically involve mass spectrometry or spectroscopy methods, including the following specific techniques: inductively coupled plasma mass spectrometry (ICP-MS) [33], time of flight and nano secondary ion mass spectrometry (SIMS) [34, 35], thermal ionization mass spectrometry (TIMS) [33], gamma spectroscopy [36], laser induced breakdown spectroscopy (LIBS), laser absorption spectroscopy (LAS), and laser induced fluorescence spectroscopy (LIFS) [37]. However, these techniques are not capable of providingsub-nanometer scale spatial resolution required to probe fine-scale microstructural features. Atom probe tomography (APT) is a 3D nanoscale compositional characterization method uniquely suited to analyze both composition and isotopic abundances in various material systems, with highspatial resolution [38-40]. Our work shows the capability to obtain nanoscale, spatially resolved quantitative $^{234,235,238}$U isotopic mapping, which has not been demonstrated before in anthropogenic U-bearing materials. We selected metallic U-10Mo nuclear fuels with two different enrichment values for demonstrating the capability of sub-nanometer scale spatially resolved, quantiative analysis of U enrichment [41-43].

## Results

### 3D elemental distribution across precipitate-matrix interfaces

In both depleted U-10Mo (DU-10Mo) and LEU-10Mo alloys, the main microstructural features are uranium carbide (UC) inclusions and the surrounding γ-UMo matrix [41]. Representative images of both DU-10Mo and LEU-10Mo microstructures are shown in Figure 1 (with additional micrographs provided in Supplementary Information, Figure S1, which provide more examples of UC morphologies). The UC phase is distributed throughout the γ-UMo matrix in both alloys, and was found to have two distinct morphologies in LEU-10Mo: (1) fine (approximately 500 nm in width), and (2) coarse (approximately 3-5 µm in width). Both UC morphologies, in addition to the γ-UMo matrix phase, were analyzed via APT in order to determine U enrichment and elemental distribution in both phases, and across γ-UMo matrix/UC interfaces.

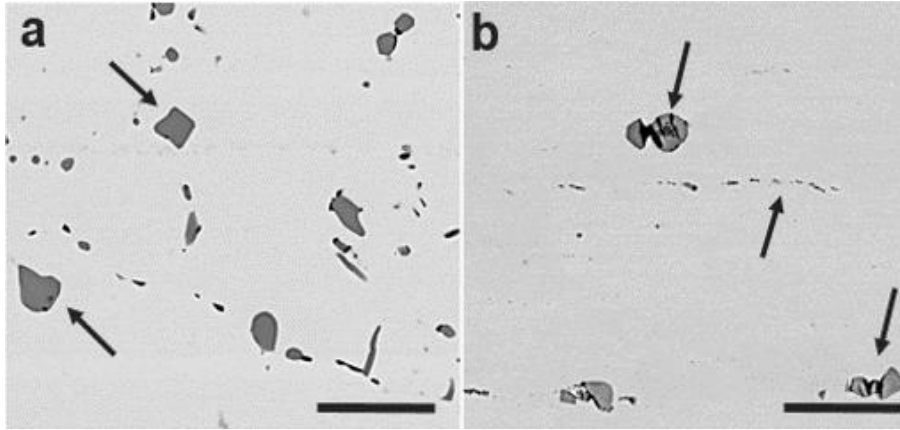

**Figure 1: Microstructure depleted and low enriched U-10Mo alloys:** Back scattered SEM images of microstructural features in the U-Mo nuclear fuel analyzed via APT in (a) DU-10Mo and (b) LEU-10Mo. Two distinct carbide morphologies are visible in these micrographs: coarse and fine. The scale bar in each micrograph is 20 µm in length.

3D distributions of major alloying elements (U, Mo) and impurity elements (Si, C) from DU-10Mo and LEU-10Mo alloys are given in Figures 2 and 3 respectively with corresponding mass spectra for UC and γ-UMo phases. Mass spectra were normalized to the peak with the maximum number of counts, which corresponds to the $^{238}U^{3+}$ peak in all data sets collected as part of this work. Normalized mass spectra were generated using manual ranging criteria. For each phase, $^{238}U^{3+}$ peak is shown in the portion of the mass spectra between mass-to-charge-state ratios (Daltons or Da) of 76 and 84. For DU-10Mo, $^{235}U^{3+}$ and $^{238}U^{3+}$ peaks were detected, and in the LEU-10Mo alloy, the $^{234}U^{3+}$ peak was also visible in the spectra, where $^{234}U$ is known to be a decay product of $^{238}U$. In depleted U, $^{234}U$ accounts for only 0.001% of all U isotopes (10 parts per million or ppm) which is approaching the detection limit for APT [38]. These results highlight the capability of APT in resolving each U isotopic peak individually which also can be spatially resolved in the 3D reconstruction.

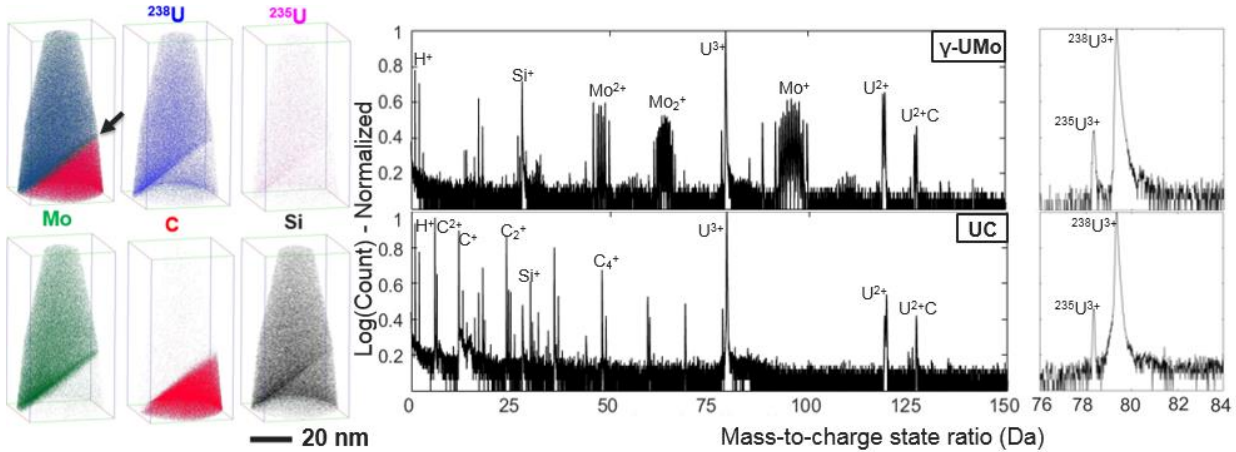

**Figure 2: APT results of depleted U-10Mo alloy**: 3D element distribution maps of U isotopes ($^{235,238}U$), Mo, C, and Si across a DU-10Mo/UC interface (indicated by the black arrow). Full mass spectra (0-150 Da) for matrix and UC phases are also provided, with the specific region of the mass spectra between 76-84 Da also shown to provide additional detail on the $U^{3+}$ peaks analyzed.

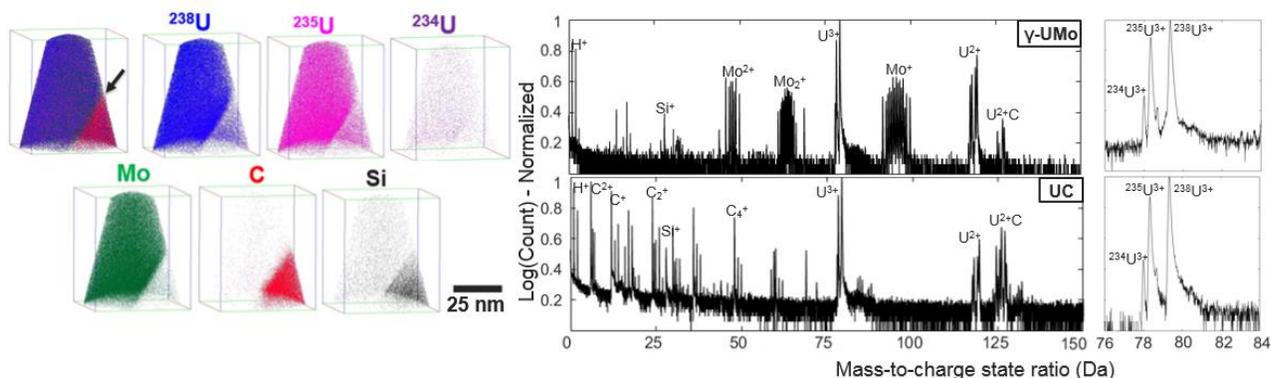

**Figure 3: APT results of low enriched U-10Mo alloy:** 3D element distribution maps of U isotopes ($^{234,235,238}$U), Mo, C, and Si across a LEU-10Mo/UC interface (indicated by the black arrow). Full mass spectra (0-150 Da) for matrix and UC phases are also provided, with the specific region of the mass spectra between 76-84 Da also shown to provide additional detail on the $U^{3+}$ peaks analyzed.

U and Mo were observed as the major species in DU-10Mo and LEU-10Mo samples, where $^{235}$U and $^{238}$U were found to field evaporate as doubly ($U^{2+}$) and triply ($U^{3+}$) charged ions. Complex ions composed of U and H (UH, or uranium hydride) were also present in mass spectra of both alloys. UH peaks were detected at approximately 118.18 and 119.66 Da (corresponding to the $U^{2+}$ charge state for $^{235}$U and $^{238}$U respecively) in LEU-10Mo. In DU-10Mo, however, only the peak at 119.66 Da was observed. The mass spectra for $U^{2+}$ for both alloys analyzed are provided in Supplementary Information, Figure S2, and show all detected UH peaks.

The change in concentration of U and Mo, and other impurity elements C and Si were quantified using a one dimensional (1D) concentration profile across the γ-UMo matrix/UC interface (Figure 4). Peak deconvolution was performed to determine total U, Mo, C, and Si contributions from both elemental and complex peaks present in the mass-to-charge spectrum, a process further detailed in the Materials and Methods section. U and Mo concentrations are consistent with alloy composition of U alloyed with 10 wt% Mo (approximately 22 at% Mo) for the γ-UMo matrix phase. In the γ-UMo matrix, C and Si impurity elements were also detected at concentrations of less than 3 at%. UC composition measured by APT ranges from 51-54 at% U and 35-39 at% C. In the UC phase, concentration of impurity elements (Si,H,O,Ni,Al) account for the balance. Analysis of impurity elements including Si, H, O, Ni, Al was performed, and the concentration of these elements was plotted across the matrix/carbide interface and are reported in Supplementary Information, Figure S3. Concentration of Si is less than 3 at%, O, Ni, and Al are all less than 2 at% in samples analyzed, whereas H concentration is more significant, and was found in concentrations of 4-20 at%. The H concentration can be attributed to the formation of UH when U is in the presence of H [17]. Sources of H leading to the formation of UH could be water used in the bulk metallographic polishing procedure, FIB prepareation of APT needle specimens and/or from residual H background from the APT analysis chamber.

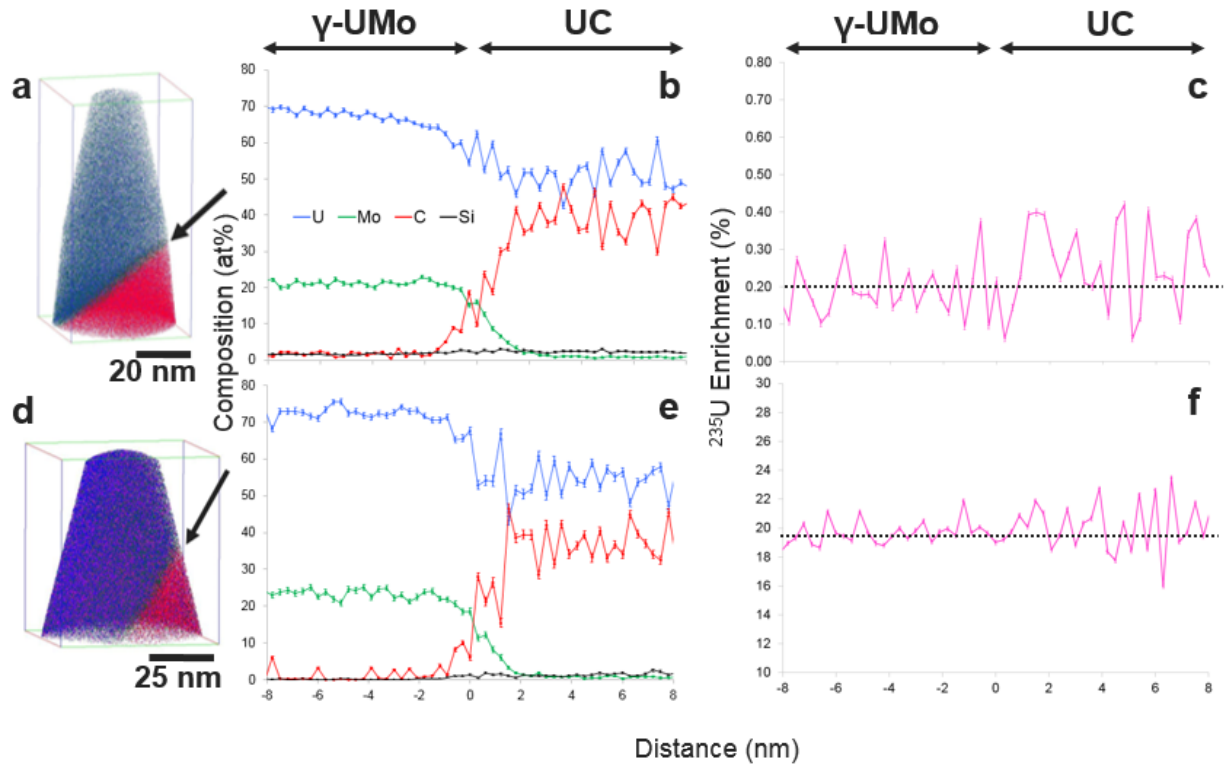

**Figure 4: Compositional and isotopic profiles of DUMo and LEUMo.**
(a) 3D element distribution map that contains a γ-UMo/UC interface in a DU-10Mo alloy, (b) corresponding compositional profile across the matrix/UC interface, (c) map of $^{235}$U% relative to all U isotopes (enrichment) across the DU-10Mo γ-UMo /UC interface. Sub-figures (d), (e), and (f) present the same type of data provided in (a),(b), and (c), but for a LEU-10Mo alloy.

## Quantification of U enrichment of individual phases

U isotope abundances in UC (fine and coarse morphologies) and γ-UMo phases were quantified from APT data, and the results are summarized in Table 1. U isotopic enrichment was estimated based on the following equation:

$$U_{enrichment} = \frac{U_i\%}{U_{total}\%}$$

Where $U_i\%$ correspond to the atom % of a particular U isotope and $U_{total}\%$ corresponds to the total uranium atom % estimated from APT results. Sample volumes from which U isotopic abundances were calculated ranged from 8,000-250,000 nm3, which corresponds to volumes with dimensions of approximately 20 nm x 20 nm x 20 nm to 50 nm × 50 nm × 100 nm. U ion counts from the $U^{3+}$ elemental peaks accounted for the majority of U ions collected, thus isotopic ratios were calculated using this peak count. Percentage of U elemental and complex species are reported in Supplementary Information, Table S1, and a comparison of calculated U enrichment from different charge state peaks are provided in Table S2. From the APT results we didn't observe any U isotopic bias when it came to preferentially evaporating as as molecular ions. Thus calculating isotopic abundances based on the major elemental peak ($U^{3+}$) provided the most consistent and accurate approach for analysis of isotopic abundances.

**Table 1: U isotopic abundances measured by APT:** Tabulated results are from different phases from both depleted and low-enriched U-Mo alloys. All data in this table is reported as percent U isotope relative to all U with percent error. Error was calculated by propagating point counting error, and is subsequently detailed in the Materials and Methods section.

| Alloy | Phase | % $^{235}$U | % $^{238}$U | % $^{234}$U |
|---|---|---|---|---|
| **DU-10Mo** | γ-UMo matrix | 0.19 ± 0.001 | 99.81 ± 0.060 | 0.0003 ± 0.0001 |
| | | 0.19 ± 0.005 | 99.81 ± 0.208 | 0.0020 ± 0.0005 |
| | UC - coarse | 0.21 ± 0.017 | 99.79 ± 0.676 | 0.0007 ± 0.0010 |
| | | 0.20 ± 0.016 | 99.79 ± 0.673 | 0.0080 ± 0.0033 |
| | | 0.21 ± 0.005 | 99.79 ± 0.210 | 0.0007 ± 0.0003 |
| **LEU-10Mo** | γ-UMo matrix | 19.47 ± 0.185 | 80.34 ± 0.527 | 0.189 ± 0.015 |
| | | 19.26 ± 0.078 | 80.51 ± 0.222 | 0.226 ± 0.007 |
| | UC - coarse | 20.44 ± 0.051 | 79.34 ± 0.138 | 0.218 ± 0.004 |
| | | 20.66 ± 0.029 | 79.11 ± 0.078 | 0.233 ± 0.003 |
| | | 20.17 ± 0.034 | 79.61 ± 0.093 | 0.219 ± 0.003 |
| | UC - fine | 20.80 ± 0.544 | 78.97 ± 1.623 | 0.229 ± 0.053 |
| | | 20.00 ± 0.042 | 79.79 ± 1.075 | 0.211 ± 0.003 |
| | | 20.04 ± 0.051 | 79.74 ± 0.141 | 0.226 ± 0.005 |

## Analysis of U enrichment across a γ-UMo grain boundary

Grain boundaries in metallic nuclear fuels are another critical interface with significant role to play when it comes to interacting with irradiation induced defects. Grain boundaries directly influence the radiation damage recovery and defect accumulation, such as fission gas bubble nucleation and growth in the irradiated fuel plate [25]. Hence as a proof-of-principle demonstration, APT was additionally used to analyze variation of isotopic enrichment across the γ-UMo grain boundary in DU-10Mo alloy. The APT reconstruction across a γ-UMo grain boundary in DU-10Mo is given in Figure 5 along with the compositional profile showing concentration of major alloying elements (U, Mo), impurity elements (Si, C, Ni, Al, Mn), and $^{235}$U enrichment. It can be seen that there is clear solute segregation along the grain boundary, consistent with our previous work [41]. Interestingly there is a narrow band of increase in the $^{235}$U enrichment specifically at the grain boundary (highlighted by the dashed black line in figure 3(c)) where locally at one spot the enrichment increased to as high as 0.5, while the overall enrichment in either grain remained consistent with the nominal value of 0.2% $^{235}$U . This highlights the unique capability of APT in not only accurately probing the compositional segregation across grain boundaries in metallic nuclear fuel but also its capability in obtaining accurate understanding of local variation in $^{235}$U enrichment across grain boundaries.

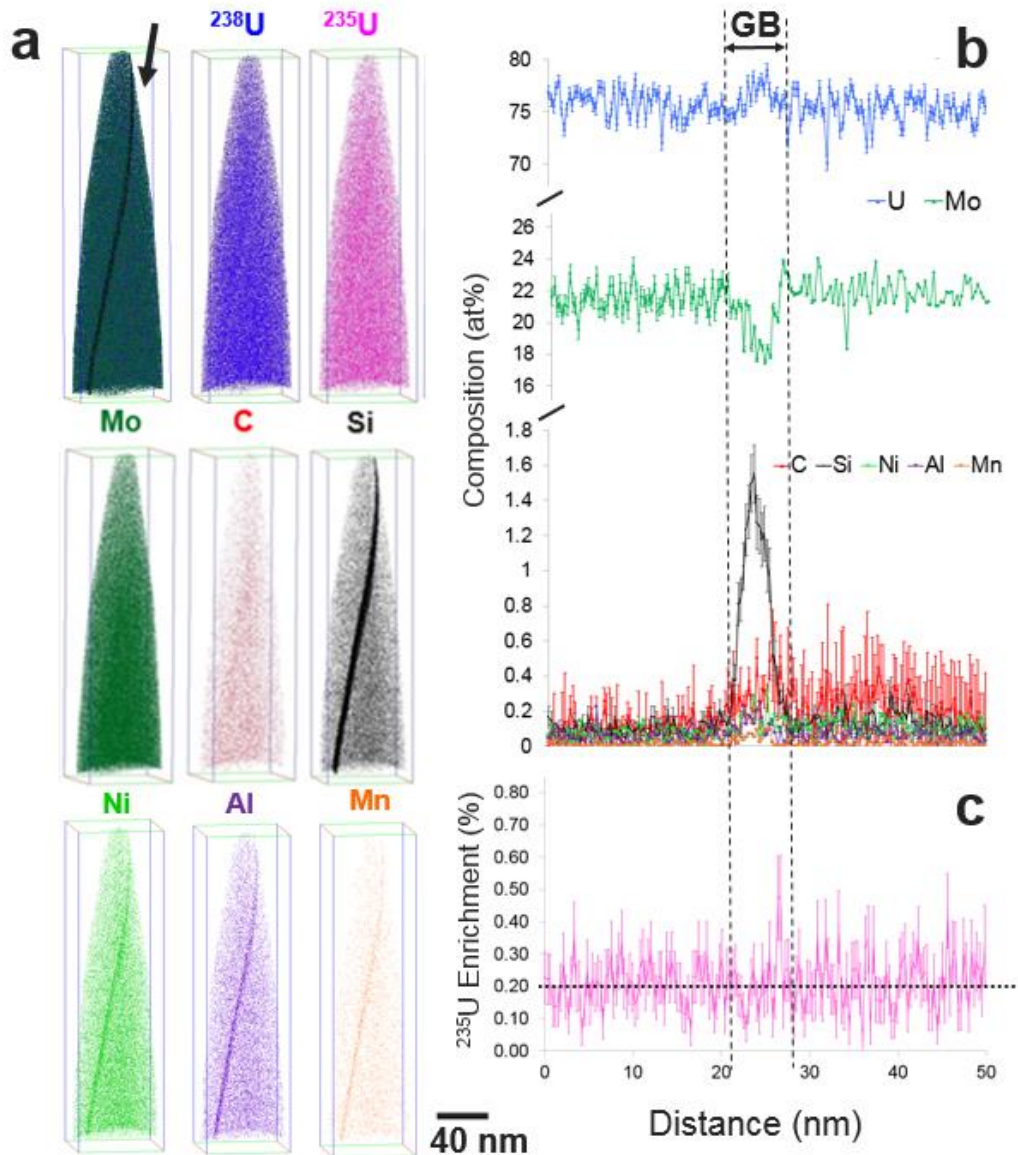

**Figure 5: Compositional and isotopic profiles of a DUMo grain boundary**, where (a) provides element distribution maps for major alloying and impurity elements, (b) corresponding quantitative analysis of composition across the grain boundary is plotted (concentration in at% versus distance), via a 1D concentration profile, and (c) $^{235}$U enrichment across the grain boundary.

## Discussion

Characterization of UC and γ-UMo matrix phases in DU and LEU alloyed with Mo was performed to demonstrate the capability of APT for quantitative measurement of U isotopes in actinide-bearing materials with varying nominal enrichments. The purpose of this effort was to gain insight into the homogeneity of $^{235}$U across different microstructural features, which has implications in fuel swelling and thus stability of nuclear fuel in a reactor operating environment.

Calculated $^{235}$U isotopic abundances for both DU-10Mo and LEU-10Mo (summarized in Table 1) indicate that U isotopic abundances are comparable between matrix and carbide phases, with minor fluctuations below 1% of U enrichment reported. This result suggests that fuel fabrication processes generate a microstructure in which $^{235}$U is relatively homogenously distributed between carbide and matrix phases. Our results indicate that the analyzed carbides likely formed during the melting and casting processes and are not retained from HEU or DU feedstock materials which are melted together and casted to form the LEU-10Mo alloy. The work presented here demonstrates that it is possible to probe U isotope abundances in fine microstructural features, and across interfaces in nuclear fuels to establish improved microstructure-processing relationships. Our results allowed us to gain insight into the sample processing history, which is a concept that could now be applied to forensics and environmental remediation studies. Additionally, the result that $^{235}$U is nominally homogenously distributed at the nanoscale between the γ-UMo matrix and UC phases can simplify fuel performance modelling since a homogenous isotopic enrichment can be assigned to describe the fuel performance in reactor.

Impurity elements (C, Si) were measured via APT, and some local increase in Si concentration was observed along the UC/ γ-UMo interface in DU-10Mo. A higher concentration of Si was observed in the UC phase of the LEU-10Mo alloy. From the U isotope abundances given in Table 1, it is clear that APT can measure the concentration of not only major isotopes of $^{235}$U and $^{238}$U, but also $^{234}$U (an indirect decay product of $^{238}$U). The APT reconstructions shown in Figures 2 and 3 from DU-10Mo and LEU-10Mo, respectively, reveal the ability to spatially resolve the isotope enrichment up to a spatial resolution of 0.2 nm in x, y and z directions. APT is also uniquely suited for analysis of enrichment across grain boundaries in nuclear fuels, which is a challenging task for most of the bulk isotopic analysis methods routinely used. Given this demonstration, our approach can now be utilized to analyze small particulates from nuclear accidents or at different stages of the nuclear fuel cycle to help analyze U enrichment and identify material origin, age, or processing history. This study opens up opportunities to utilize APT as a method of choice for a variety of other fields ranging from earth and planetary sciences to bioremediation, toxicology and environmental monitoring.

In summary, by using site specific sample preparation and laser assisted APT analysis, we demonstrate the unique capability to perform spatially resolved nanoscale mapping of isotopic enrichment and impurity elements in U-Mo fuels with two different $^{235}$U enrichments. Based on these results, we believe APT has a strong prospect for use in a wide range of characterization efforts of actinide-bearing materials in which isotopic analysis in small volumes (and of specific microstructural features) is critical. Our work has implications for multiple disciples that could benefit from precise measurement of U isotope enrichment and fractionation to better understand sample origin or processing history. Methods and experimental protocols using APT presented here translate well to analysis of U enrichment in many other systems (geologic materials, metallic fuels, and glass fabricated as a nuclear waste form, for example). Thus, APT is promising as an impactful and versatile tool for measurement of isotopic abundances in actinide-bearing materials, at length scales previously unexplored by other more widely used mass spectrometry and spectroscopy methods.

## Materials and Methods
### Material Fabrication

DU-10Mo and LEU-10Mo fuels analyzed in this work were fabricated at the Y12 National Security Complex at Oak Ridge National Laboratory. DU-10Mo samples were fabricated by melting DU with Mo and casting in a graphite mold. The cast DU-10Mo was then homogenized at 900 C for 48 hours in an inert gas (argon) atmosphere. After the homogenization annealing treatment, samples were cooled by turning off the furnace and forcing argon gas into the chamber to assist in sample cooling. A cooling rate of 25 C per minute was achieved from 900 C to 650 C, then 3 C per minute from 500 C to 350 C. LEU-10Mo samples were homogenized at 900 C for 144 hours in the same atmosphere and furnace as described previously. The LEU-10Mo was subjected to subsequent thermo-mechanical processing treatments (hot and cold rolling) to form the final fuel foil. Additional sample fabrication details are reported in References [47, 19, 17]

## FIB-based sample preparation and APT analysis

APT needles were prepared from standard metallographicaly polished bulk samples [32]. Site-specific FIB lift-outs and annular milling was performed according to methods described in References [8].

A CAMECA LEAP 4000X HR system equipped with a 355 nm wavelength UV laser was used for all APT data collection with the following user-selected parameters: 100 pJ laser energy, 100 kHz pulse frequency, 45 K specimen temperature, and 0.005 atoms/pulse detection rate. Data sets analyzed and reported here ranged in size between 1.4 and 34.5 million ions, and either contain a single phase (UC or γ-UMo matrix) or capture a UC/γ-UMo interface. All data sets were reconstructed and analyzed using the Interactive Visualization and Analysis Software (IVAS), version 3.8.2. For data sets containing a single phase, bulk composition analysis was completed, and for those data sets that contained an interface, composition in each phase was determined using spherical regions of interest (ROIs) with diameters of 20-30 nm.

To examine composition across the interface, a cylindrical ROI with a diameter of 20 nm was positioned perpendicular to the planar interface, and a one dimensional (1D) composition profile along the z-axis with a 0.3 nm bin width was constructed. For all composition analysis, ionic and background corrected data was used.

In order to quantify Mo and C concentrations accurately from raw data, peak deconvolution was performed according to methods presented in Reference [44]. Peak deconvolution was required to distinguish between $^{96}Mo^{2+}$ and C4 species that both have a mass-to-charge state ratio of approximately 48 Da, and therefore cannot be ranged separately. The number of $^{96}Mo^{2+}$ ions was estimated from the number of $^{98}Mo^{2+}$ ions, assuming both $^{96}Mo^{2+}$ and $^{98}Mo^{2+}$ are present in U-Mo at naturally abundant levels, similar to the process reported in Reference [40]. Equation 1 was used to determine the $^{96}Mo^{2+}$ ion count:

$$n_{^{96}Mo^{2+}} = n_{^{98}Mo^{2+}} \times \frac{0.1668}{0.2413} \qquad (1)$$

where $n_{^{96}Mo^{2+}}$ is the estimated ion count for the $^{96}Mo^{2+}$, and $n_{^{98}Mo^{+2}}$ is the ion count for the $^{98}Mo^{2+}$ peak. The two fractions in Equation 1 (0.1668 and 0.2413) are the natural isotope abundances of $^{96}Mo$ and $^{98}Mo$, respectively. Since the entire $^{96}Mo^{2+}$ peak was ranged as a single species but not all counts correspond to $^{96}Mo^{2+}$, the calculated value of $n_{^{96}Mo^{2+}}$ was subtracted from the total ion counts corresponding to the ranged peak to obtain the ion count for the C4 species.

In order to obtain an accurate estimation of C concentration in the volume analyzed, we used the $^{13}C$ method detailed in Reference [45]. This approach involves multiplying the $^{13}C$ background corrected ion count by 92.5 to determine a more accurate value for the $^{12}C$ ion count. The 92.5 multiplier relies on the assumption that C exists in natural abundance levels.

Peak deconvolution was performed via methods described above to analyze the total U, Mo, C, and Si concentration across the interface. Error reported in tables and composition profiles represent propagation of point counting error, defined as:

$$E = \sqrt{\frac{C_i(1-C_i)}{N}} \quad (2)$$

where $C_i$ is $i$ number of solute ions, $N$ is the total number of ions in a bin, and $x_i$ is the count for a specific element. $C_i$ is defined in mathematical form as:

$$C_i = \frac{x_i}{N} \quad (3)$$

Error reported as a percentage is simply the value calculated from Equation 2 multiplied by 100. Error associated with U enrichment was calculated with Equations 4-5, where $E_{Total,U}$ (Equation 4) is the total error associated with all U isotope counts, $U_{enrichment}$ and $E_{U_{enrichment}}$ (Equation 5) is the counting error associated with a U isotope enrichment value.

$$E_{Total,U} = \sqrt{(E_{234_U})^2 + (E_{235_U})^2 + (E_{238_U})^2} \quad (4)$$

$$E_{U_{enrichment}} = U_i \times \sqrt{\left(\frac{E_{U_i}}{U_i}\right)^2 + \left(\frac{E_{U_{total}}}{U_{total,conc}}\right)^2} \quad (5)$$

**ACKNOWLEDGEMENTS**

This work was supported by the U.S. Department of Energy (DOE), National Nuclear Security Administration under Contract DE-AC05-76RL01830. A portion of this research was performed using facilities at the Environmental Molecular Sciences Laboratory, a national scientific user facility sponsored by the DOE's Office of Biological and Environmental Research and located at Pacific Northwest National Laboratory (PNNL). The authors thank Mark Rhodes of PNNL for performing bulk metallographic sample preparation, and all the other staff directly or indirectly associated with producing the results featured in this publication.

**AUTHOR CONTRIBUTIONS**

A.D. conceptualized and directed all work presented here. E.K. and A. D. performed site-specific sample preparation for analysis of interfaces and conducted the APT experiments. E.K. and A.D. analyzed all APT data. C.L. and V.J. coordinated the fabrication and metallographic preparation of DU-10Mo and LEU-10Mo samples used in this work. D.B. contributed to manuscript writing, and discussions about results and its significance to in reactor performance of nuclear fuels. All authors contributed to the discussion of results and manuscript preparation.

**Additional Information**
**Competing interests:** The author(s) declare no competing interests.

# References

[1] J. R. Bargar *et al.*, "Uranium redox transition pathways in acetate-amended sediments," (in en), *PNAS,* vol. 110, no. 12, pp. 4506–4511, March 2013 2013.

[2] G. A. Brennecka and M. Wadhwa, "Uranium isotope compositions of the basaltic angrite meteorites and the chronological implications for the early Solar System," (in en), *PNAS,* vol. 109, no. 24, pp. 9299–9303, June 2012 2012.

[3] J. Hiess, D. J. Condon, N. McLean, and S. R. Noble, "238U/235U Systematics in Terrestrial Uranium-Bearing Minerals," *Science,* vol. 335, no. 6076, pp. 1610–1614, 2012 2012.

[4] C. H. Stirling, "Keeping Time with Earth's Heaviest Element," *Science,* vol. 335, no. 6076,

pp. 1585–1586, 2012 2012.

[5] M. Stylo *et al.*, "Uranium isotopes fingerprint biotic reduction," (in en), *PNAS,* vol. 112, no. 18, pp. 5619–5624, May 2015 2015.

[6] H.-D. Sues, "Dating the origin of dinosaurs," (in en), *PNAS,* vol. 113, no. 3, pp. 480–481, January 2016 2016.

[7] G. A. Wagner *et al.*, "Radiometric dating of the type-site for Homo heidelbergensis at Mauer, Germany," (in en), *PNAS,* vol. 107, no. 46, pp. 19726–19730, November 2010 2010.

[8] R. E. Wood, C. Barroso-Ruíz, M. Caparrós, J. F. J. Pardo, B. G. Santos, and T. F. G. Higham, "Radiocarbon dating casts doubt on the late chronology of the Middle to Upper Palaeolithic transition in southern Iberia," (in en), *PNAS,* vol. 110, no. 8, pp. 2781–2786, February 2013 2013.

[9] R. J. M. Konings, T. Wiss, and O. Beneš. (2015). *Predicting material release during a nuclear reactor accident*    [Comments and Opinion].

[10] X. Wang, Z. Xu, A. Soulami, X. Hu, C. Lavender, and V. Joshi, "Modeling Early-Stage Processes of U-10 Wt.%Mo Alloy Using Integrated Computational Materials Engineering Concepts," *JOM,* vol. 69, no. 12, pp. 2532–2537, December 2017 2017.

[11] S. Takahashi, Ed. *Radiation Monitoring and Dose Estimation of the Fukushima Nuclear Accident*. Springer Nature, 2017.

[12] S. F. Boulyga *et al.*, "Determination of 236U/238U isotope ratio in contaminated environmental samples using different ICP-MS instruments," *Journal of Analytical ATomic Spectrometry,* vol. 17, pp. 958-964, 2002 2002.

[13] J. N. Christensen, P. E. Dresel, M. E. Conrad, G. W. Patton, and D. J. DePaolo, "Isotopic Tracking of Hanford 300 Area Derived Uranium in the Columbia River," *Environmental science & technology,* vol. 44, no. 23, pp. 8855–8862, 2010 2010.

[14] A. J. Fahey, C. J. Zeissler, D. E. Newbury, J. Davis, and R. M. Lindstrom, "Postdetonation nuclear debris for attribution," *Proceedings of the National Academy of Sciences,* vol. 107, no. 47, pp. 20207–20212, 2010 2010.

[15] S. K. Hanson *et al.*, "Measurements of extinct fission products in nuclear bomb debris: Determination of the yield of the Trinity nuclear test 70 y later," (in en), *PNAS,* vol. 113, no. 29, pp. 8104–8108, July 2016 2016.

[16] R. C. Marin, J. E. S. Sarkis, and M. R. L. Nascimento, "The use of LA-SF-ICP-MS for nuclear forensics purposes: uranium isotope ratio analysis," *Journal of Radioanalytical and Nuclear Chemistry,* vol. 295, no. 1, pp. 99–104, January 2013 2013.

[17] T. L. Martin *et al.*, "Atomic-scale Studies of Uranium Oxidation and Corrosion by Water Vapour," (in en), *Scientific Reports,* vol. 6, p. 25618, July 2016 2016.

[18] S. Van den berghe and P. Lemoine, "Review of 15 years of high-density low-enriched UMo dispersion fuel development for research reactors in Europe," *Nuclear Engineering and Technology,* vol. 46, no. 2, pp. 125–146, April 2014 2014.

[19] J. R. Vavrek, B. S. Henderson, and A. Danagoulian, "Experimental demonstration of an isotope-sensitive warhead verification technique using nuclear resonance fluorescence," (in en), *PNAS,* vol. 115, no. 17, pp. 4363–4368, April 2018 2018.

[20] J. D. Ward *et al.*, "Identifying anthropogenic uranium compounds using soft X-ray near-edge absorption spectroscopy," *Spectrochimica Acta Part B: Atomic Spectroscopy,* vol.

127, pp. 20–27, January 2017 2017.

[21] M. R. B. I. Andrew Conant, Anna Erickson, "Sensitivity and Uncertainty Analysis of Plutonium and Cesium Isotopes in Modeling of BR3 Reactor Spent Fuel," *Nuclear Technology,* vol. 197, no. 1, pp. 12-19, 2017 2017.

[22] E. Kuhn, D. Fischer, and M. Ryjinski, "Environmental sampling for IAEA safeguards: A five year review," International Atomic Energy Agency, Vienna, AustriaIAEA-SM–367, 2001 2001.

[23] L. J. H. James J. Duderstadt, *Nuclear Reactor Analysis*. Wiley, 1976.

[24] S. Hu, D. E. Burkes, C. A. Lavender, D. J. Senor, W. Setyawan, and Z. Xu, "Formation mechanism of gas bubble superlattice in UMo metal fuels: Phase-field modeling investigation," *Journal of Nuclear Materials,* vol. 479, pp. 202 - 215, 2016 2016.

[25] M. K. Meyer *et al.*, "Low-temperature irradiation behavior of uranium–molybdenum alloy dispersion fuel," *Journal of Nuclear Materials,* vol. 304, no. 2, pp. 221–236, August 2002 2002.

[26] Y. S. Kim and G. Hofman, "Fission product induced swelling of U-Mo alloy fuel," *Journal of Nuclear Materials,* vol. 419, pp. 291–301, December 2011 2011.

[27] S. Neogy, A. Laik, M. T. Saify, S. K. Jha, D. Srivastava, and G. K. Dey, "Microstructural Evolution of the Interdiffusion Zone between U-9 Wt Pct Mo Fuel Alloy and Zr-1 Wt Pct Nb Cladding Alloy Upon Annealing," (in en), *Metall and Mat Trans A,* vol. 48, no. 6, pp. 2819–2833, June 2017 2017.

[28] J. L. Snelgrove, G. L. Hofman, M. K. Meyer, C. L. Trybus, and T. C. Wiencek, "Development of very-high-density low-enriched-uranium fuels1Work supported by the US Department of Energy, Office of Nonproliferation and National Security, under contract No. W-31-109-ENG-38.1," *Nuclear Engineering and Design,* vol. 178, no. 1, pp. 119–126, December 1997 1997.

[29] M. Ugajin, A. Itoh, M. Akabori, N. Ooka, and Y. Nakakura, "Irradiation behavior of high uranium-density alloys in the plate fuels," *Journal of Nuclear Materials,* vol. 254, no. 1, pp. 78 - 83, March 1998 1998.

[30] *Research Reactors: Safe Management and Effective Utilization* (Proceedings Series - International Atomic Energy Agency). Vienna, Austria: IAEA, 2015.

[31] S. Van den Berghe, W. Van Renterghem, and A. Leenaers, "Transmission electron microscopy investigation of irradiated U–7wt%Mo dispersion fuel," *Journal of Nuclear Materials,* vol. 375, no. 3, pp. 340-346, 2008/04/30/ 2008.

[32] H. J. Ryu, Y. S. Kim, and G. L. Hofman, "Amorphization of the interaction products in U–Mo/Al dispersion fuel during irradiation," *Journal of Nuclear Materials,* vol. 385, no. 3, pp. 623-628, 2009/04/15/ 2009.

[33] J. S. Becker, "Chapter 13 - Inorganic Mass Spectrometry of Radionuclides," in *Handbook of Radioactivity Analysis (Third Edition)*, M. F. L'Annunziata, Ed. Amsterdam: Academic Press, 2012, pp. 833-870.

[34] D. Huang, X. Hua, G.-L. Xiu, Y.-J. Zheng, X.-Y. Yu, and Y.-T. Long, "Secondary ion mass spectrometry: The application in the analysis of atmospheric particulate matter," *Analytica Chimica Acta,* vol. 989, pp. 1-14, 2017/10/09/ 2017.

[35] A. Krein *et al.*, "Imaging Chemical Patches on Near-surface Atmospheric Dust Particles with NanoSIMS 50 to Identify Material Sources," *Water, Air, & Soil Pollution: Focus,* vol.

8, no. 5, pp. 495-503, 2008/12/01 2008.
[36] C. T. Nguyen and J. Zsigrai, "Basic characterization of highly enriched uranium by gamma spectrometry," *Nuclear Instruments and Methods in Physics Research Section B: Beam Interactions with Materials and Atoms,* vol. 246, no. 2, pp. 417-424, 2006/05/01/ 2006.
[37] S. S. Harilal, B. E. Brumfield, N. L. LaHaye, K. C. Hartig, and M. C. Phillips, "Optical spectroscopy of laser-produced plasmas for standoff isotopic analysis," *Applied Physics Reviews,* vol. 5, no. 2, p. 021301, 2018/06/01 2018.
[38] A. Devaraj *et al.*, "Three-dimensional nanoscale characterisation of materials by atom probe tomography," *International Materials Reviews,* vol. 63, no. 2, pp. 68–101, February 2018 2018.
[39] S. Piazolo *et al.*, "Deformation-induced trace element redistribution in zircon revealed using atom probe tomography," (in en), *Nature Communications,* vol. 7, p. 10490, February 2016 2016.
[40] L. White *et al.*, *Atomic-scale age resolution of planetary events*. 2017.
[41] A. Devaraj *et al.*, "Grain boundary engineering to control the discontinuous precipitation in multicomponent U10Mo alloy," *Acta Materialia,* vol. 151, pp. 181–190, June 2018 2018.
[42] S. Jana *et al.*, "Kinetics of cellular transformation and competing precipitation mechanisms during sub-eutectoid annealing of U10Mo alloys," *Journal of Alloys and Compounds,* vol. 723, pp. 757–771, November 2017 2017.
[43] A. Devaraj *et al.*, "Phase transformation of metastable discontinuous precipitation products to equilibrium phases in U10Mo alloys," *Scripta Materialia,* vol. 156, pp. 70-74, 2018/11/01/ 2018.
[44] S. D. Taylor, J. Liu, B. W. Arey, D. K. Schreiber, D. E. Perea, and K. M. Rosso, "Resolving Iron(II) Sorption and Oxidative Growth on Hematite (001) Using Atom Probe Tomography," *J. Phys. Chem. C,* vol. 122, no. 7, pp. 3903–3914, February 2018 2018.
[45] M. Thuvander *et al.*, "Quantitative atom probe analysis of carbides," *Ultramicroscopy,* vol. 111, no. 6, pp. 604–608, May 2011 2011.